\documentclass[%
aip,
 amsmath,amssymb,
 reprint,%
]{revtex4-1}

\usepackage{graphicx}
\usepackage{dcolumn}
\usepackage{bm}

\usepackage[utf8]{inputenc}
\usepackage[T1]{fontenc}
\usepackage{mathptmx}
\usepackage{etoolbox}
\usepackage{hyperref}

\makeatletter
\def\@email#1#2{%
 \endgroup
 \patchcmd{\titleblock@produce}
  {\frontmatter@RRAPformat}
  {\frontmatter@RRAPformat{\produce@RRAP{*#1\href{mailto:#2}{#2}}}\frontmatter@RRAPformat}
  {}{}
}%
\makeatother
\begin{document}

\preprint{AIP/123-QED}

\title[Intelligent Control of MeV Electrons and Protons]{Towards Intelligent Control of MeV 
Electrons and Protons from kHz Repetition Rate Ultra-Intense Laser 
Interactions}
\author{Nathaniel Tamminga}
\affiliation{Department of Physics, The Ohio State University, Columbus, OH 43210, USA}
\email{tamminga.2@osu.edu}
\author{Scott Feister}
\affiliation{Department of Computer Science, California State University Channel Islands, Camarillo, CA 93012, USA}
\author{Kyle D. Frische}
\affiliation{Air Force Institute of Technology, Wright-Patterson AFB, OH 45433, USA}
\author{Ronak Desai}
\affiliation{Department of Physics, The Ohio State University, Columbus, OH 43210, USA}
\author{Joseph Snyder}
\affiliation{Department of Mathematical and Physical Sciences, Miami University, Hamilton, OH 45011, USA}
\author{John J. Felice}
\affiliation{Department of Physics, The Ohio State University, Columbus, OH 43210, USA}
\author{Joseph R. Smith}
\affiliation{Department of Physics, Marietta College, Marietta, OH 43750, USA}
\author{Chris Orban}
\affiliation{Department of Physics, The Ohio State University, Columbus, OH 43210, USA}
\author{Enam A. Chowdhury}
\affiliation{Department of Materials Science, The Ohio State University, Columbus, OH 43210, USA}
\author{Michael L. Dexter}
\affiliation{Air Force Institute of Technology, Wright-Patterson AFB, OH 45433, USA}
\author{Anil K. Patnaik}
\affiliation{Air Force Institute of Technology, Wright-Patterson AFB, OH 45433, USA}

\date{\today}

\begin{abstract}
Ultra-intense laser-matter interactions are often difficult to predict from first principles because of the complexity of plasma processes and the many degrees of freedom relating to the laser and target parameters. An important approach to controlling and optimizing ultra-intense laser interactions involves gathering large data sets and using this data to train statistical and machine learning models. In this paper we describe experimental efforts to accelerate electrons and protons to $\sim$MeV energies with this goal in mind. These experiments involve a 1~kHz repetition rate ultra-intense laser system with $\sim$10~mJ per shot, a peak intensity near $5 \times 10^{18}$~W/cm$^{2}$, and a ``liquid leaf" target. Improvements to the data acquisition capabilities of this laser system greatly aided this investigation. Generally, we find that the trained models were very effective for controlling the numbers of MeV electrons ejected. The models were less successful at shifting the energy range of ejected electrons. Simultaneous control of the numbers of $\sim$MeV electrons and the energy range will be the subject of future experimentation using this platform.
\end{abstract}


\maketitle

\section{\label{sec:intro}Introduction}
As discussed in two recent reviews \cite{dopp_etal2023, Anirudh_etal2022}, machine learning (ML) methods have increasingly become a useful tool in experimental plasma physics.

While there are a number of papers where machine learning methods have been used to optimize and control electron beams via laser wakefield \cite{He_etal2015, Lin_etal2021, Jalas_etal2021, Ye_etal2022, Dann_etal2019}, there has been less experimental work in optimizing and controlling MeV electron sources from laser interactions with denser targets (exceptions include \cite{Mariscal_etal2024}). This is often referred to as the direct laser acceleration (DLA) regime and these interactions typically produce a broad energy spectrum of electrons. Another research opportunity in ultra-intense laser science is the application of ML methods to controlling  MeV proton acceleration \cite{Noaman_etal2018,   Ma_etal2021}. A perspectives article by \citet{Palmer2018} argues that a compact and tunable proton source like this would be desirable for a variety of scientific, defense and industrial purposes. Although energetic electrons have different practical applications than energetic protons, the electron dynamics produce the electric fields that accelerate the protons. Measurements of MeV electrons can therefore illuminate efforts to produce MeV protons, and so we consider both electrons and protons in this paper.

A key technology to achieve precision control over laser based proton and electron sources is high repetition rate ultra-intense laser systems \cite{Hooker2013}. When these systems are coupled to high throughput data acquisition systems \cite{Heuer_etal2022,Feister_etal2023}, large data sets can be created that can be used to train ML models. These trained ML models can aid in determining the laser and target parameters that would produce electron or proton energy spectra with desired properties. In the literature, some studies characterize laser repetition rates of 0.2~Hz as being frequent enough to be considered ``high repetition rate". But, for our purposes, because many ML models typically involve thousands of free parameters or more that need to be constrained in order for the ML model to be useful, the repetition rate needs to be much larger. We show results from an ultra-intense laser system operating at a 1 kHz rate with advanced diagnostics that take advantage of that capability. Previously mentioned efforts that use ML to control and optimize laser wakefield likewise tend to use repetition rates in the kHz range (e.g. \cite{He_etal2015}).

The experimental work presented here is part of a wider effort to optimize and control multiple sources of secondary radiation from the 10~mJ class, 1~kHz repetition rate laser system at Wright Patterson AFB. This system has previously been used to produce MeV protons \cite{Morrison_etal2018}, MeV electrons \cite{Feister_etal2017}, x-rays \cite{Morrison_etal2015}, and recently neutrons \cite{Knight_etal2024}. However, none of these experiments utilize ML in efforts to control the secondary radiation as explored in this paper.

The work presented here benefits from theoretical efforts in \citet{Desai_etal2024} in which synthetic data sets were generated for a 10~mJ class ultra-intense laser system where different machine learning models were tested on that data. That study was focused on proton acceleration but many of the lessons learned were applied to the experimental work presented here.

In Sec.~\ref{sec:exp} we describe the ultra-intense laser system, target setup and data acquisition system. In Sec.~\ref{sec:design} we describe the design of our experimental campaign. In Sec.~\ref{sec:results} we describe the results from the experimental campaign which includes a parameter scan that was used to train regression models on the dependence of MeV electron generation on the laser-target parameters. This campaign also includes an effort to use the trained models to optimize and control the MeV electron source. In Sec.~\ref{sec:discuss} we discuss the results and in Sec.~\ref{sec:concl} we conclude. We also provide in Appendix~\ref{ap:delay} an additional experimental investigation where the pre-pulse time delay is varied.

\section{\label{sec:exp}Experimental Setup}
\subsection{\label{sec:exp:laser}Laser System and Target Generation}
\begin{figure}
    \centering
    \includegraphics[width=\linewidth]{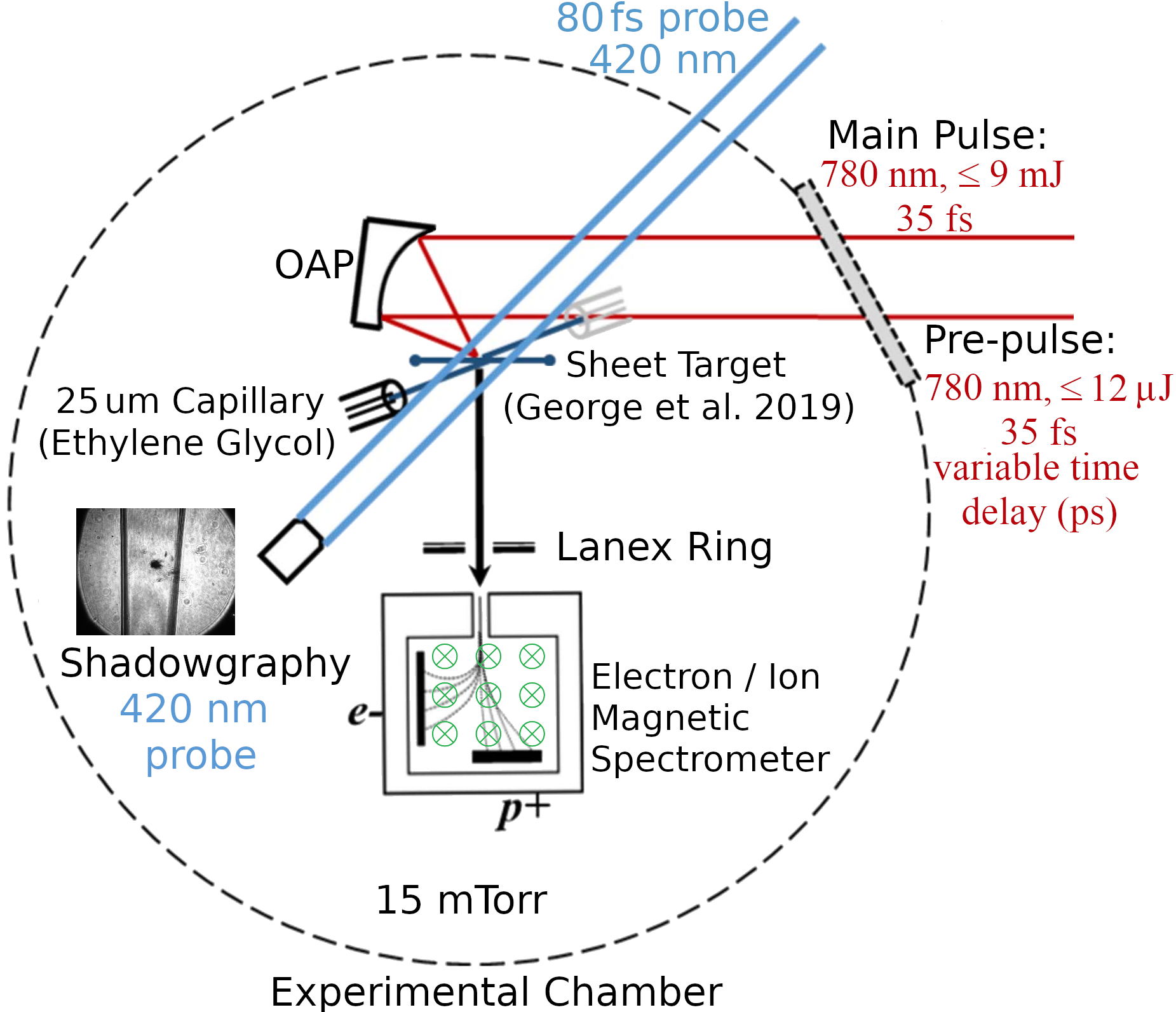}
    \caption{A schematic of the experimental chamber, the targetry system and the laser pulses entering the chamber at a kHz repetition rate. The liquid target system also operates at a kHz as described in \citet{George_etal2019}. A magnetic spectrometer provides information on the number and energies of electrons and ions ejected from the laser-target interaction.}
    \label{fig:exp}
\end{figure}

The experimental setup is shown in Fig.~\ref{fig:exp}. Three laser pulses enter the experimental chamber: the ``main" laser pulse, an intentional pre-pulse, with energy, time delay, and pulse duration control, and a probe pulse. The main pulse provides up to $\sim$9~mJ of energy with a $\sim$35~fs (FWHM) duration at 780~nm wavelength. The intentional pre-pulse with up to 12~$\mu$J of energy also enters the chamber with 35~fs duration and 780~nm wavelength. Both the main pulse and the pre-pulse are focused onto the target using a \textit{f}/1 gold-coated off axis parabolic mirror. At peak focus the main pulse spot size is close to 1.8~$\mu$m FWHM. The main pulse maximum intensity is near $5 \times 10^{18}$~W/cm$^{2}$. A waveplate and two thin-film polarizers (TFP) combination allows precision control of the main pulse energy up to a factor of ten. The magnetic electron/ion spectrometer seen in Fig.~\ref{fig:exp} is discussed in Sec.~\ref{sec:exp:epics}.

The pre-pulse is produced when the main laser pulse passes through a beam splitter that diverts a small amount of laser energy to a different chain of optics. Along this path is a delay stage where a set of mirrors on a translation stage is used to set the difference in timing between the main pulse and pre-pulse. The pre-pulse is recombined with the main pulse, via a TFP, and propagates collinearly to the OAP and to the target. The pre-pulse and main pulse are orthognally polarized. The delay stage provides sub-picosecond control of the time delay. Similar to the main pulse, the pre-pulse can be attenuated in a precise and controlled way.

Shadowgraphy of the target is obtained with a probe laser pulse. As discussed in \citet{Feister_etal2014}, the probe beam is temporally synchronized to the main laser and frequency doubled and frequency shifted to a central wavelength of 420~nm. The probe pulse duration is 80~fs and the pulse energy is $\sim$60~$\mu$J.

The main pulse and pre-pulse are focused onto a free-flowing ethylene-glycol sheet. A liquid sheet is generated by colliding two liquid jets in a method previously described \cite{George_etal2019}. The laser interaction destroys the sheet but the sheet will quickly re-form \cite{George_etal2019}.

\subsection{\label{sec:exp:epics}Data Acquisition System}
As discussed in \citet{Feister_etal2023}, state of the art high-repetition-rate ultra-intense laser facilities are moving to a ``distributed network control system" \cite{Ge_etal2017} model where instruments are wired to a network via gigabit ethernet and data is sent in packets via a network protocol. As discussed in \citet{Feister_etal2023}, a framework called the Experimental Physics Industrial Control System (EPICS) is designed to interface with control devices (pumps, sensors, valves, etc.) on this distributed network control system. With the release of EPICS v7 and the pVAccess network protocol, this framework is very well suited for high repetition rate laser experiments.

\begin{figure*}
    \centering
    \includegraphics[width=\linewidth]{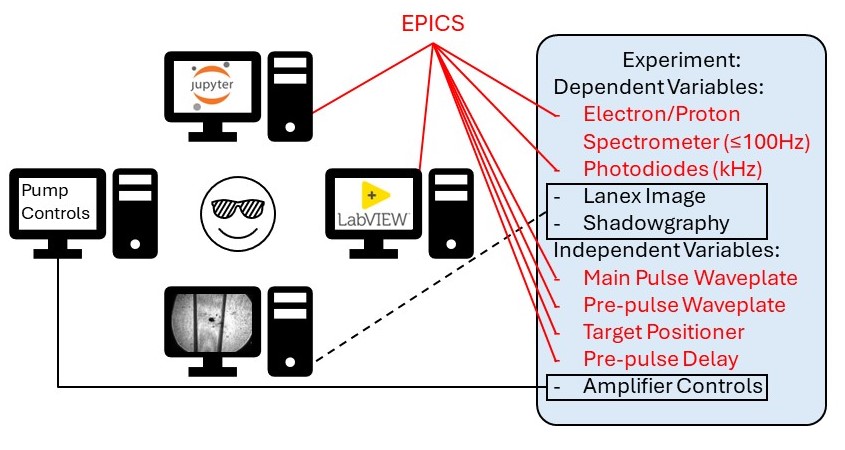}
    \caption{Implementation of EPICS-based devices working alongside existing, non-EPICS, controls during each experimental run. Components in red are EPICS-based and components in black utilize preexisting control software.}
    \label{fig:EPICS}
\end{figure*}

Our laser system is notable in that the control and data acquisition is a hybrid system that includes both EPICS-based controls and direct-connection controls. Fig.~\ref{fig:EPICS} shows a high level overview of this system. Originally the laser system operated purely with controls based on direct connection between the user interface computer and the device. This interface was limiting to our data acquisition needs. Our system was running with LabView in a synchronous manner and parameters could not be changed until all the data was saved. Firmware and software changes along with the introduction of EPICS allowed for the use of a common shot count across all data devices, which allows for asynchronous operation and data collection with the ability to combine the data for analysis with the common shot count. This improvement allows parameters to vary during an experiment while we know what changes correspond to different electron and ion spectra. We migrated important instruments and target and laser controls to EPICS. In this way, our laser system is an interesting example of modernizing a data acquisition and laser control system without completely re-designing it from the ground up. This hybrid approach provides two different ways to control and interface with the EPICS devices. This control is possible because LabView can be configured to send commands over EPICS and receive the data, or python scripts running independently of LabView can execute these commands. The former is useful for adjusting experimental conditions by hand while the latter is useful for allowing machine learning algorithms to direct the laser system to new parameters to explore as we will demonstrate in the next section.

The instruments and controls that were migrated to EPICS include Thorlabs DET10A and DET100A diodes that measure the main pulse and pre-pulse energy at a kHz rate, a magnetic electron/ion spectrometer that was used in earlier experiments (e.g. \cite{Feister_etal2017,Morrison_etal2018}), a Newport XPS-Q8 8-axis universal controller and driver, a Newport CONEX-URS50BCC motorized rotation stage, and Lin Engineering's R256 and R365 stepper motor drivers. With the exception of the spectrometer, all control  and data acquisition devices are connected to commercially available BeagleBone Black microcontrollers which are responsible for either data acquisition or serving as an EPICS interface between the instrument and the network. For simplicity, all EPICS devices were set up using the StreamDevice module from EPICS v7 \cite{streamdevice_2024}. This method was selected because StreamDevice is a generic communication interface for serial line, IEEE-488, and telnet-like TCP/IP communications, which allowed us to use it for multiple devices and greatly reduced the complexity and development time when integrating EPICS into our existing lab setup.

Of particular importance to this work is the electron/ion spectrometer which is connected to the EPICS servers through two virtual machines (one for electron, one for ion) which also serve as data recorders. As shown in Fig.~\ref{fig:exp}, this device includes two line CCDs so that electrons are magnetically deflected to one line CCD and positively charged ions are magnetically deflected in the other direction to a different line CCD. Because the laser is interacting with ethylene glycol, one expects protons as well as Carbon and Oxygen ions to be ejected from the interaction region. Unfortunately, because there is magnetic but not electric field deflection, we cannot distinguish between these ion species. As in an earlier paper published by our group \cite{Morrison_etal2018}, we assume that protons represent the majority of the signal, and we map the pixel positions to the kinetic energy in MeV based on this assumption even though there could be some Carbon and Oxygen ions that arrive at these same positions on the CCD with a different energy than we are quoting for protons. 

Importantly, our electron/ion spectrometer is well suited for the high repetition rate regime. In prior work from our group\cite{Feister_etal2017,Morrison_etal2018}, this device operated with a repetition rate of 10~Hz. We were able to increase the repetition rate to 100~Hz for the first time, which is the limit of the line CCD on the device. In terms of the data acquisition, the client interfaces primarily consist of LabVIEW and CS-Studio (Phoebus) interfaces. CS-Studio is a control system software allows users to build their own control system displays and controls. LabVIEW interacts with EPICS using the CALab user library \cite{calab_2024}. This capability was especially useful because most of our old control interfaces were LabVIEW-based controls, allowing the front-facing control panel to look identical with only the back panel code changing. The data from the laser diodes, the electron/ion spectrometer and other measureables such as the target position, are saved in hdf5 format.

\section{\label{sec:design}The Design of An Experimental Campaign}

\renewcommand{\arraystretch}{1.25}
\begin{table}
    \centering
    \begin{tabular}{|p{2.84cm}|p{1.1cm}|p{1.1cm}|p{1.7cm}|}
         \hline
         Explored Parameter & Max & Min & Scan Period\\
         \hline \hline
         Main Pulse Energy & 8.4~mJ & $\lesssim$1~mJ & 80s\\
         \hline
         Pre-Pulse Energy & 12~$\mu$J & $\lesssim$1~$\mu$J & 104s\\
         \hline
         Target Distance from Optimum Focal Plane & +30~$\mu$m & -30~$\mu$m & 700s\\
         \hline
    \end{tabular}
    \caption{A description of the experimental parameters, the range they were scanned through, and the period of the parameter scan.}
    \label{tab:exp_runs}
\end{table}
\renewcommand{\arraystretch}{2}

As mentioned in \ref{sec:exp:laser}, the liquid target can only form for approximately 45 minutes before needing to pause the experiment to refill the tanks. We refer to one $\sim$~45~minute data collection period as an experimental run. Data was collected in two runs. The first run was a parameter scan while the second was an optimization effort. In the parameter scan, different laser and target parameters were systematically varied in an effort to explore the parameter space. These varied parameters are shown in Table~\ref{tab:exp_runs}. In an optimization run, the information from the parameter scan is used in conjunction with different algorithms to predict the laser and target conditions that will produce a desired effect on the electron energy distribution. A primary goal is to determine which algorithm is most effective in manipulating and optimizing the electron energy distribution, which will be discussed later.

As described in Table~\ref{tab:exp_runs}, the parameter scan simultaneously varied the target position, main pulse energy and pre-pulse energy. The scan was designed so that the main pulse and pre-pulse have slightly different scanning frequencies in order to investigate many different combinations of main pulse and pre-pulse energy. The main pulse energy was varied from its highest value to its lowest value and back to the highest value over a timescale of 80~seconds. Likewise the pre-pulse energy was varied over a timescale of 104~seconds. This approach explores the parameter space of main pulse and pre-pulse energy in a relatively uniform way as will be discussed in Sec.~\ref{sec:results}.

The target position was varied more slowly. The data collection began with the target near the peak focus and then the target was moved to $30~\mu$m away from the focus in one direction and then it was moved $30~\mu$m away from the focus in the other direction. The full exploration of the target position (including the return to the initial target position) takes 700~seconds.

In the optimization run, three different regression algorithms were trained with the data from the parameter scan and then used to predict the main pulse energy, pre-pulse energy and target focal position that modifies the electron energy distribution in a desired way as discussed in the next section. The effort to migrate instruments and diagnostics to the EPICS interfaced network greatly enabled the optimization effort because EPICS commands can be sent using a python interface and there are many optimization libraries and ML methods that are written in python. 

\subsection{\label{sec:design:opt}Optimization}

The optimization run uses three different regression algorithms in an effort to modify the electron energy spectrum in a controlled way. Electron energy spectra typically involve a range of kinetic energies. In principle, there are many ways that one could try to change the energy spectrum. For simplicity, we focus on shifting the peak kinetic energy, and increasing the number of electrons. We do this by defining an objective function that the algorithms will try to optimize,
\begin{eqnarray}
     f(KE_{\rm cutoff},N_{e}) & =  & 1 + (1 - \beta) \cdot \left(1 -  \frac{\log_{10}(N_{e})}{8} \right)  \nonumber  \\
     & & + \beta \frac{|KE_{\rm cutoff} - KE_{\rm cutoff,goal}|}{KE_{\rm cutoff, goal}}  \nonumber \\ \label{eq:function}
\end{eqnarray}
where $KE_{\rm cutoff}$ is the maximum electron kinetic energy (99th percentile), $KE_{\rm cutoff,goal}$ is the desired maximum kinetic energy, $N_e$ is the total number of electron counts from the electron spectrometer and $\beta$ is a parameter that determines the relative importance of matching the maximum electron kinetic energy ($\beta = 1$) compared to increasing the number of electrons ($\beta = 0$). The spectrometer CCD has 3648 pixels with a maximum of $2^{16}$ counts per pixel so a completely saturated spectrum would have $\log_{10}(N_e) = \log_{10}(2^{16} \cdot 3648) = 8.38$. This is the reason for dividing $\log_{10} (N_e)$ by 8 in Eq.~\ref{eq:function}.

The CCDs contain 13 light shielded pixels whose average value provide an estimate of the background signal. 
To remove noise, we subtract the raw CCD signal by a factor of 115 \% of the average value of the light shielded pixels for each shot. Then, to smooth out the signal, we apply a median filter that replaces each pixel value with the inclusive median of the 15 closest pixels.

\subsection{\label{sec:design:algo}Regression Algorithms}

Three different regression algorithms were trained on the data from the parameter scan: a polynomial ``Ridge" regression, random forest regression and Gaussian Process Regression. We used the scikit-learn implementation of these algorithms \cite{scikit-learn} and we used the default hyperparameters from scikit-learn unless otherwise noted. These algorithms were then used to predict the laser independent variables that minimize Eq.~\ref{eq:function} for different values of $\beta$ and $KE_{\rm cutoff, goal}$.

We trained polynomial ridge regression models with degrees 2, 4, 6, and 8. `Ridge' refers to the regularization parameter determined through cross-validation (via RidgeCV). We trained random forest regression (RF) models with 100 or 500 estimators (trees) with a max depth of 10 (see RandomForestRegressor). We also used a gaussian process regression (GPR) with a noise level ($\alpha$) of either 2 or 10. The noise helps the GPR deal with instabilities during fitting (see GaussianProcessRegressor). In the optimization run shown in the next section, each of these algorithms require only about a minute or less on a CPU to be trained on the data from the parameter scan and predict the laser independent variables that minimize Eq.~\ref{eq:function}.

\section{\label{sec:results}Results}
\subsection{\label{sec:results:ps}Parameter scan}

\begin{figure*}
    \centering
    \includegraphics[width=\linewidth]{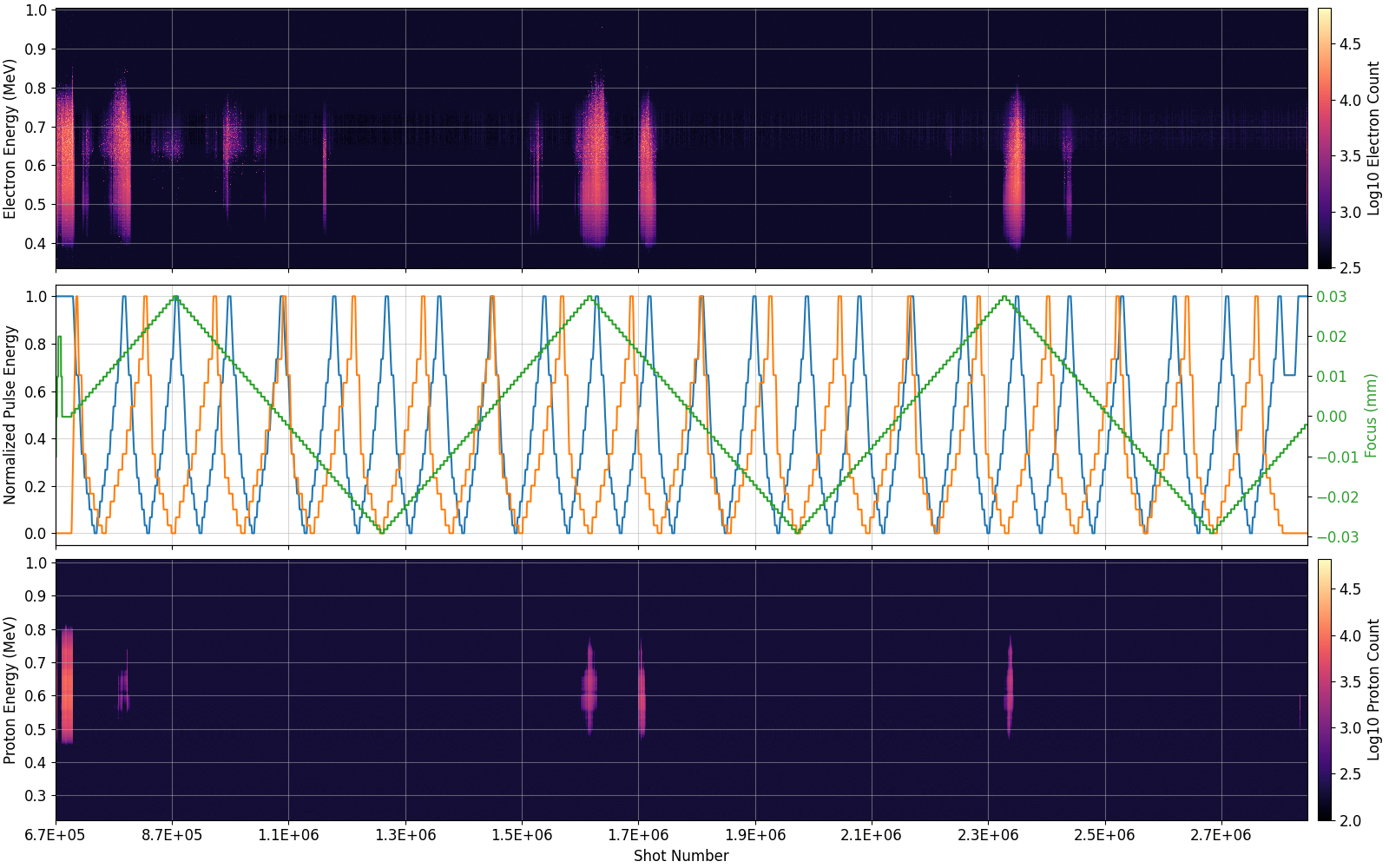}
    \caption{Results from the parameter scan. The x-axis shows the shot number with respect to the beginning of the experimental run. The upper panel shows the electron spectra over time. Center panel: A plot showing the normalized main pulse (blue), pre-pulse (orange) and target focal position (green) as these quantities were varied over time. The bottom panel shows the proton spectra over time. Both the electron and proton spectrometers were collecting at a rate of 10~Hz and are plotted logarithmically with peak value corresponding to single pixel saturation of the CCD.}
    \label{fig:focus_looping_multi}
\end{figure*}

We performed a parameter scan which lasted 35 minutes and involved roughly 2.25 million laser shots. Fig.~\ref{fig:focus_looping_multi} shows the results of the parameter scan. The center panel of Fig. \ref{fig:focus_looping_multi} shows the parameters varied while the upper and lower panels show the electron and proton spectra collected during the parameter scan. In these experimental runs, the pre-pulse time delay relative to the main pulse was fixed at 300~ps (see Appendix~\ref{ap:delay} for additional experimental data about the sensitivity to the time delay). 

\begin{figure*}
    \centering
    \includegraphics[width=0.49\linewidth]{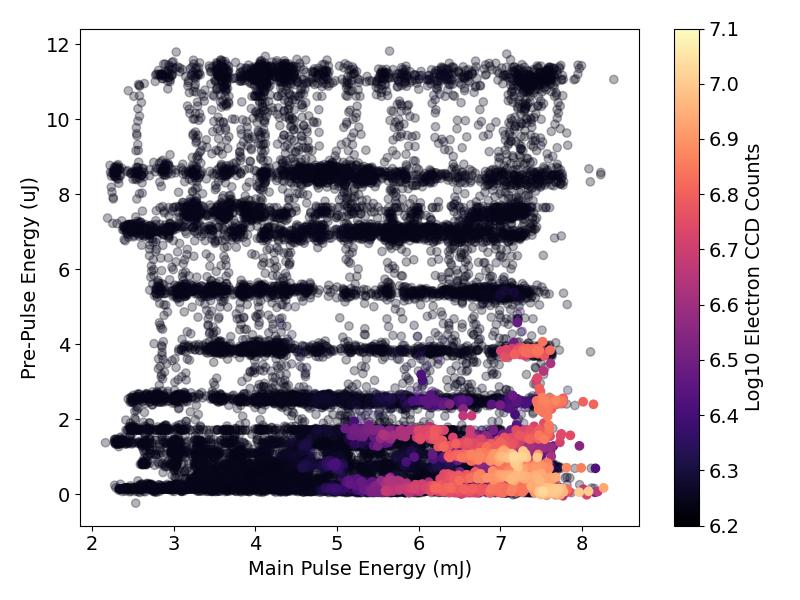}
    \includegraphics[width=0.49\linewidth]{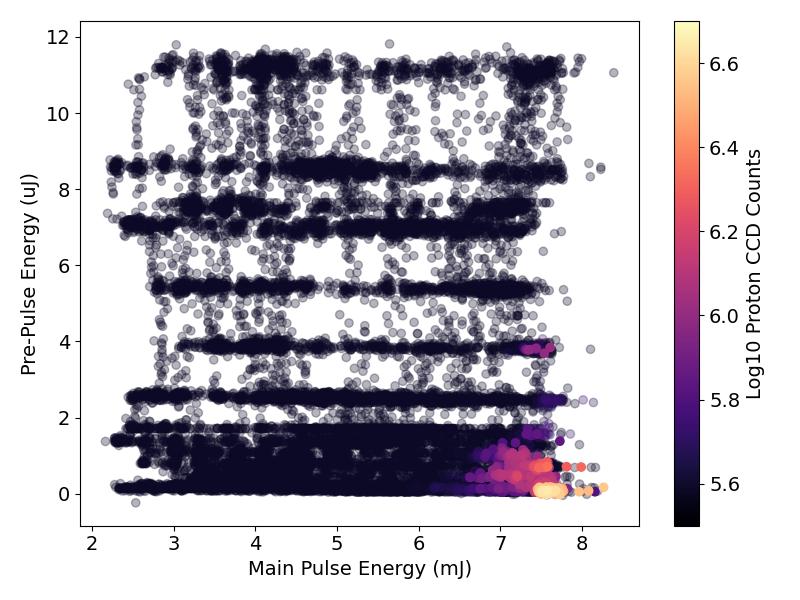}
    \vspace{-0.7cm}
    \caption{Results from the parameter scan showing the dependence of accelerated particle counts on pre-pulse and main pulse energy. The left panel shows electron number counts while the right panel shows proton number counts. The colorbars show the log of the electron or proton number counts. Note that the target position was also varied during this parameter scan which is why one can see high and low number count results even for similar combinations of main pulse and pre-pulse energy.}
    \label{fig:Param_space_scatter1}
\end{figure*}

The left and right panels of Fig.~\ref{fig:Param_space_scatter1} show the results from the electron and ion spectrometer versus pre-pulse and main pulse energy. Both plots show a clear trend for higher number counts with higher main pulse energy, as long as the pre-pulse level is low. These plots include all of the data from Fig.~\ref{fig:focus_looping_multi} which means that results from target positions both near and far from peak focus are included. In both plots in Fig.~\ref{fig:Param_space_scatter1}, lower count points are plotted underneath higher count points. Far from the peak focus, the effective intensity on target can be relatively low even if the main pulse energy is at its maximum value. Consequently, in Fig.~\ref{fig:Param_space_scatter1}, one can see low count results near high count results even when the main pulse energy is high and the pre-pulse energy is low. 

Both plots in Fig.~\ref{fig:Param_space_scatter1} have horizontal features. These features are real and they reflect the way that the pre-pulse energy was incremented in ten steps (see for example the discrete steps of the orange line in the center panel of Fig.~\ref{fig:focus_looping_multi}). The main pulse energy was also incremented in steps but the shot-to-shot variations in main pulse energy are large enough that it is difficult to notice.

\subsection{\label{sec:results:opt}Optimization}
After training regression algorithms on data gathered from the parameter scan, these trained models were used in continued experimentation in efforts to optimize the electron signal for $KE_{\rm cutoff}=0.7-1.0$~MeV and the value of $\beta$ varying from $\beta=0.0-1.0$. This ``optimization" run lasted 48 minutes and involved roughly 2.8 million laser shots. These regressions established the mapping between the dependent variables minimized through Equation \ref{eq:function} (e.g. the electron counts and cutoff energy) and the independent variables that found these conditions (e.g. main pulse waveplate, pre-pulse waveplate, and target position). In this way, specifying a value for $\beta$ and $KE_{\rm cutoff, goal}$ uniquely specifies a shot number from the parameter scan whose corresponding independent variables are repeated in the optimization run.

\begin{figure*}
    \centering
    \includegraphics[width=0.49\linewidth]{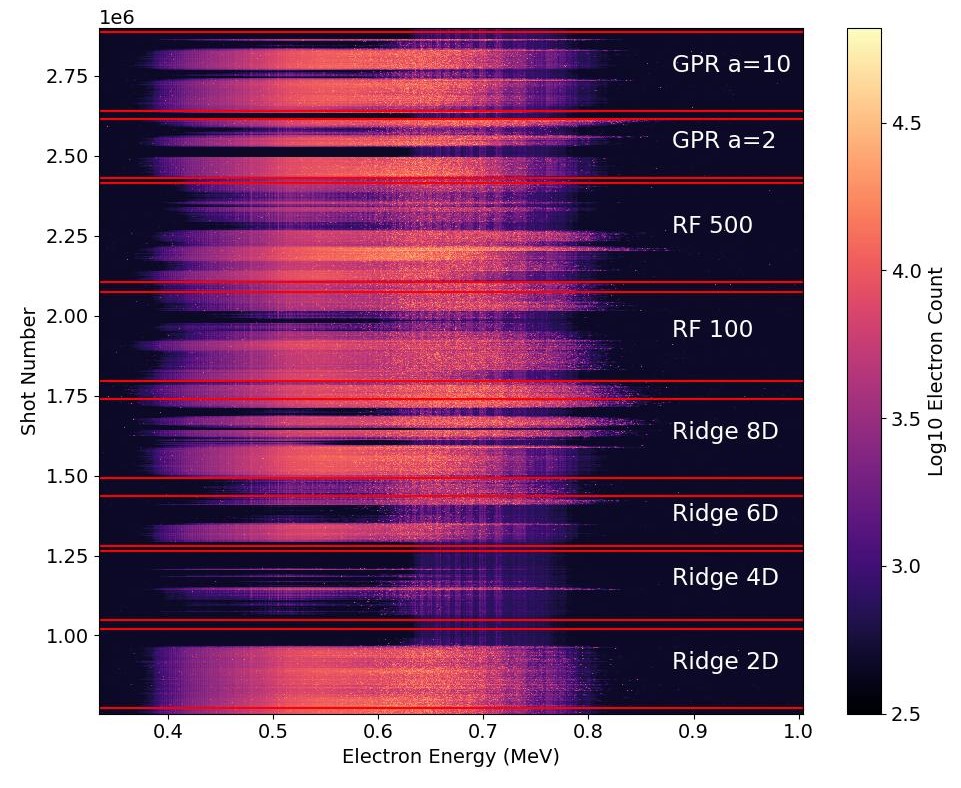}
    \includegraphics[width=0.49\linewidth]{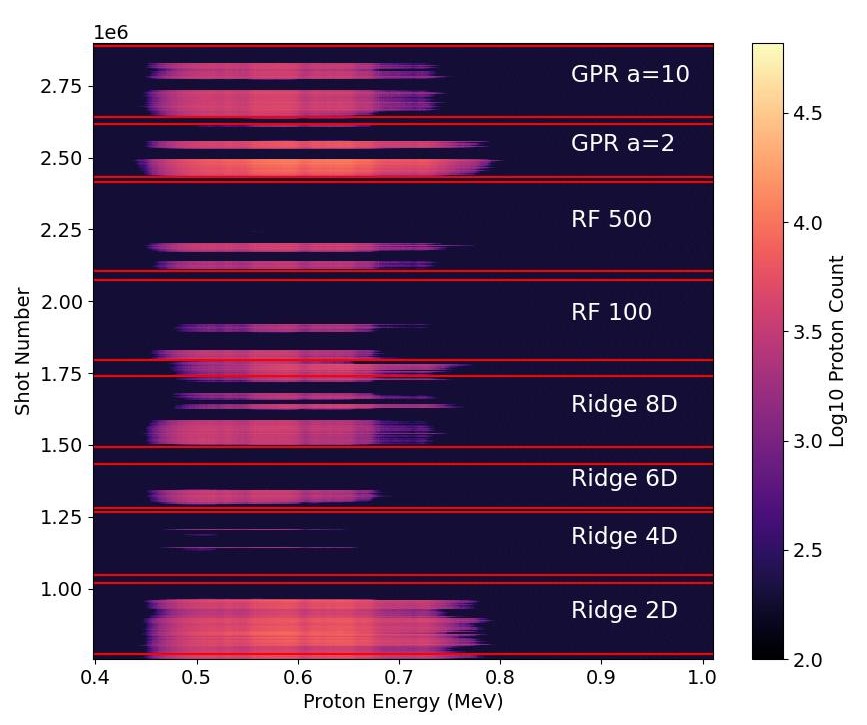}
    \vspace{-0.5cm}
    \caption{Optimization results. Electron (left) and proton (right) spectra are shown with the optimization model listed to the right-hand side. Both the electron and proton color bars are logarithmic.
    Variability in the spectra over time for each model is a consequence of exploring different predictions for the independent variables that optimize Eq.~\ref{eq:function} for different values of $KE_{\rm cutoff, goal}$ and $\beta$ as discussed in the text. Eq.~\ref{eq:function} only considers the measured electron energy spectrum but we include the proton results here because protons were often detected during this investigation.}
    \label{fig:opt_spectrum}
\end{figure*}

In total, eight models were tested which were based off of the three previously discussed regression algorithms. The raw results can be seen in Figure \ref{fig:opt_spectrum}. The objective function did not include any information from the proton spectrum, but protons were detected during the optimization run so we show these results in right panel of Figure \ref{fig:opt_spectrum}.

\begin{figure}
    \centering
    \includegraphics[width=\linewidth]{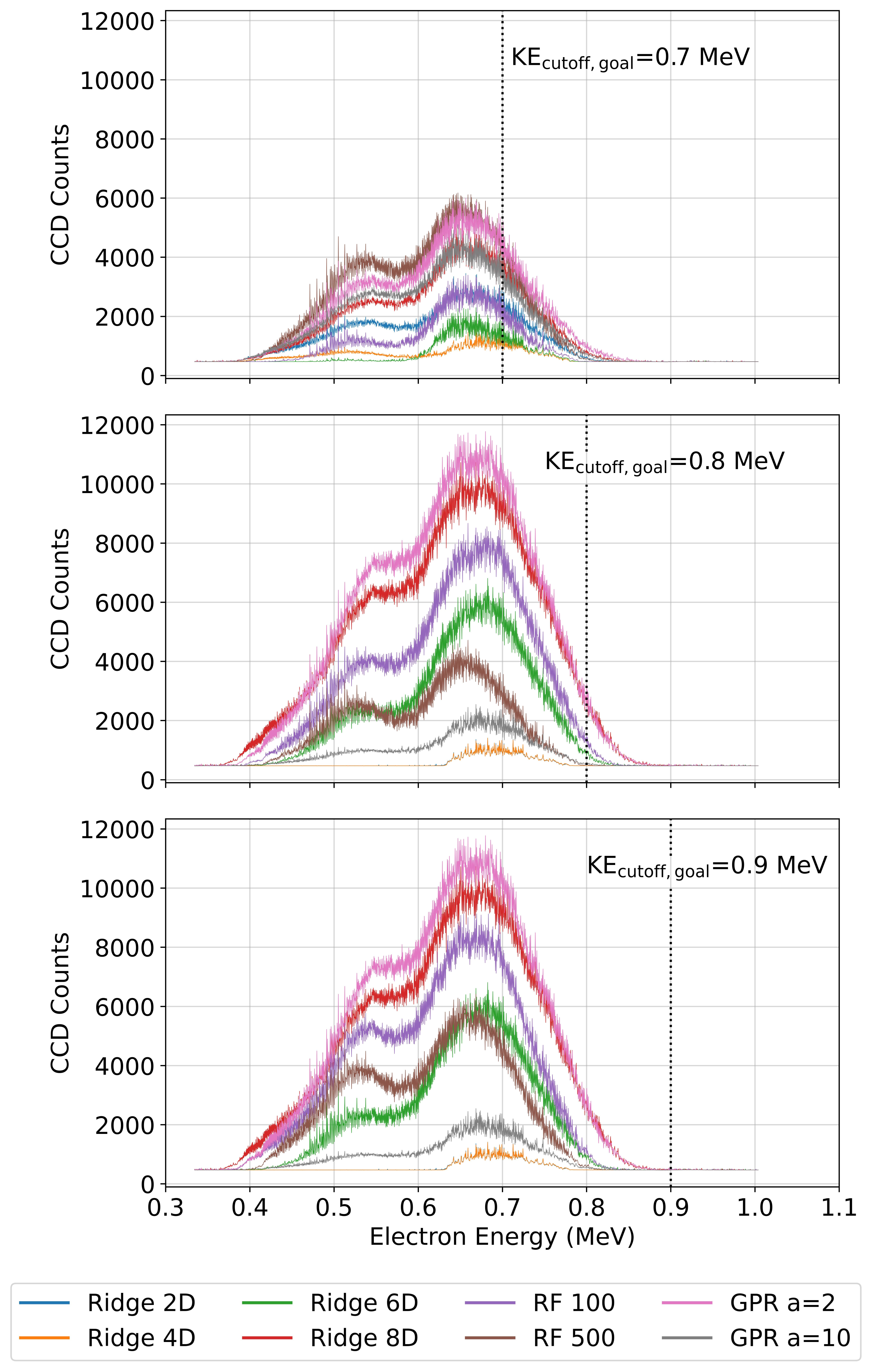}
    \caption{Electron spectra for $\beta=1$  and varying values of  $KE_{\rm cutoff}$, shown by the vertical black dashed line on each plot. There is a shift in the peak energy between $KE_{\rm cutoff}=0.7-0.8$~MeV.}
    \label{fig:spectra_energies}
\end{figure}

In Figure \ref{fig:spectra_energies}, the electron energy spectrum for each model is shown for $\beta=1$ and with different values for $KE_{\rm cutoff, goal}$ which is noted by vertical black dotted lines. Because this test used $\beta = 1$, which completely removes the total electron number counts from the optimization, this investigation acts as kind of a stress test to see how well the models can match the cutoff energy.

\begin{figure}
    \centering
    \includegraphics[width=\linewidth]{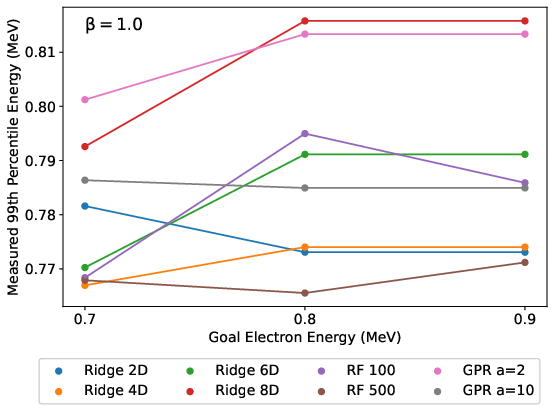}
    \caption{Shift of the $99^{th}$ percentile of the electron count for different optimization models with $\beta=1$ for varying $KE_{\rm cutoff, goal}$.}
    \label{fig:energy_shift}
\end{figure}

Figure~\ref{fig:energy_shift} summarizes results from Figure~\ref{fig:spectra_energies} by showing the $99^{th}$ percentile of the electron count versus $KE_{\rm cutoff, goal}$ for each model. We plot the $99^{th}$ percentile energy because this is the same metric that the optimization algorithm used to determine $KE_{\rm cutoff}$ in Eq~\ref{eq:function}. Many of the models increased their energy when the optimizer changed $KE_{\rm cutoff, goal}$ from 0.7~MeV to 0.8~MeV. Exceptions were the 2-dimensional ridge regression, 500 estimator random forest, and Gaussian process with $\alpha=10$. The latter two had only a small drop in their energy. The $99^{th}$ percentile electron energy did not increase when $KE_{\rm cutoff, goal}$ increased to 0.9~MeV and the 500 estimator random forest even dropped in energy. We comment on this result in the discussion section.

\begin{figure}
    \centering
    \includegraphics[width=\linewidth]{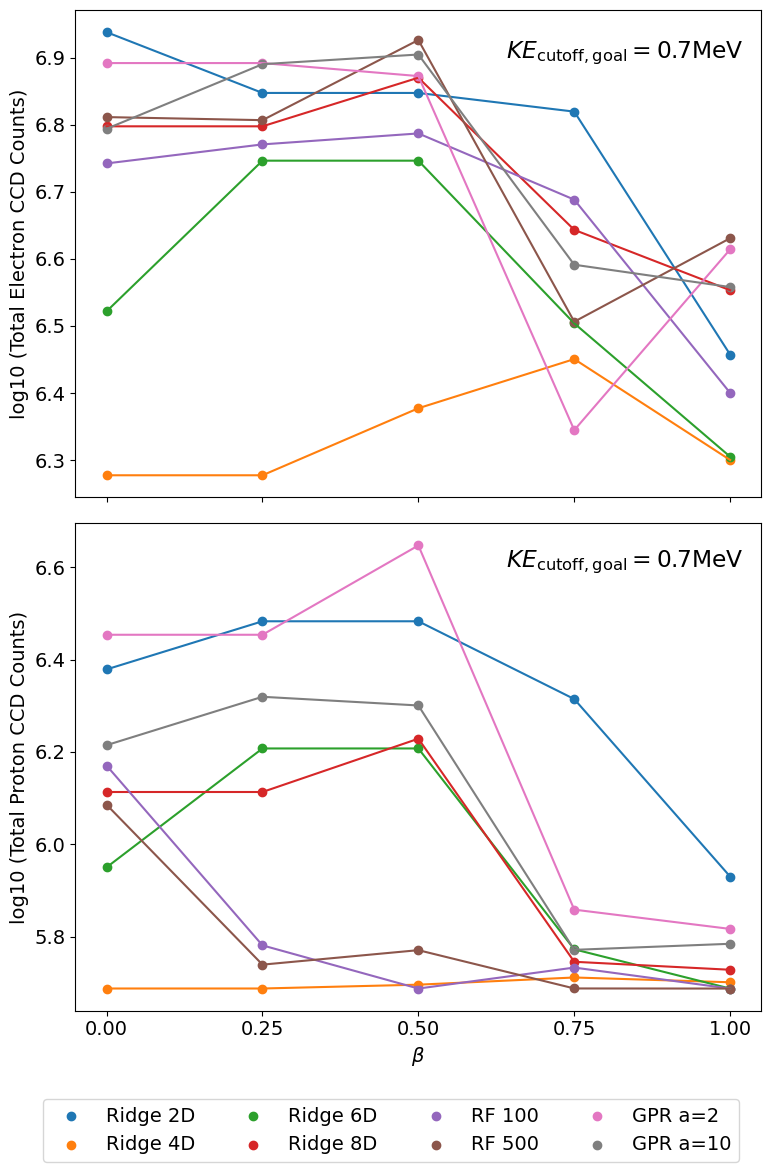}
    \caption{Total electron (upper) and proton (lower) CCD counts for different models while varying the electron $\beta$ from 0 to 1 and keeping $KE_{\rm cutoff, goal}$ fixed at 0.7~MeV. The Ridge 4D model results were affected by target instability but the other models were not.}
    \label{fig:beta_shift}
\end{figure}

Figure \ref{fig:beta_shift} shows the log of the total electron (upper) and proton (lower) CCD count for varying electron $\beta$ values with $KE_{\rm cutoff, goal}=0.7$~MeV. $KE_{\rm cutoff}=0.7$~MeV was chosen as the cutoff energy for this evaluation because each spectra has significant electron counts around 0.7~MeV. All models, except the 4-dimensional ridge regression, exhibit an increase in total electron counts as $\beta$ is varied from 1 to 0 as one would expect given the definition of Eq.~\ref{eq:function}. Likewise, the total proton counts increase with the electron counts as the electron $\beta$ is varied from 1 to 0. Instead of showing detailed data for each model, we highlight some typical results from one of the models that performed well in Figure~\ref{fig:beta_shift}. Specifically, we show the electron and proton energy spectra from the Gaussian process ($\alpha = 2$) model. One can see that as $\beta$ decreases the total electron and proton number counts increase significantly.

We note that in Figure~\ref{fig:beta_shift}, the 4-dimensional ridge regression did not decrease because there was significant target instability during that optimization.  In Figure~\ref{fig:opt_spectrum}, the 4-dimensional ridge regression model produced very little electron CCD counts during the testing of that model.  

\begin{figure}
    \centering
    \includegraphics[width=\linewidth]{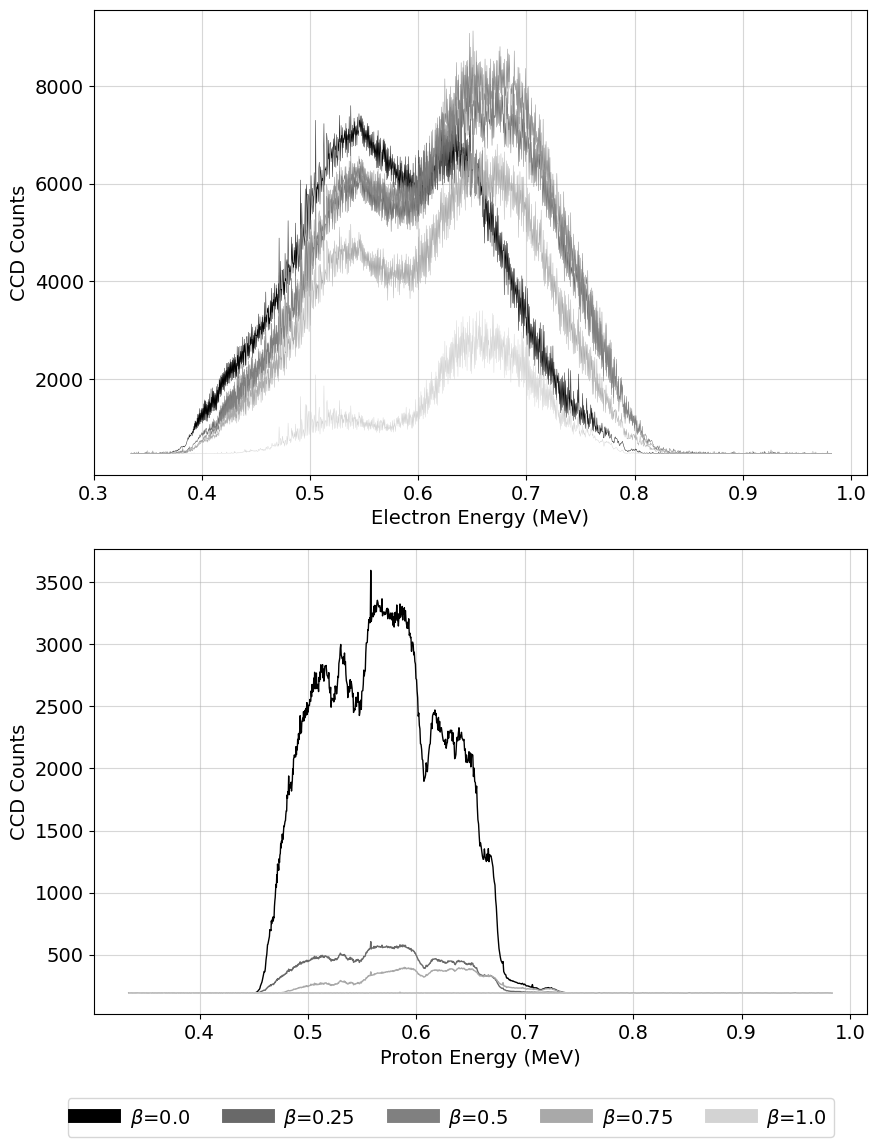}
    \caption{Electron (upper) and proton (lower) energy spectra for the random forest regression with 100 estimators. Results shown are for $\beta=0$ to $\beta=1$ while keeping $KE_{\rm cutoff, goal}$ fixed at 0.7~MeV.}
    \label{fig:beta_spectra}
\end{figure}

\section{\label{sec:discuss}Discussion}
This work was only possible due to significant improvements to the data acquisition system in our lab which included retrofitting our system with high throughput control using EPICS. This modification allowed for automated control of the parameters suggested via the optimization algorithm and data collection at a higher throughput rate. Single shot diagnostics of main pulse and pre-pulse energy were collected from over 5-million laser shots. The electron/ion spectrometer was improved to operate at 100~Hz but we collected at a rate of 10~Hz in order to increase the signal to noise. In total, over 50,000 unique electron and proton spectra were obtained for different experimental conditions, including different main pulse and pre-pulse energies, and target position. In this data, the time delay between the pre-pulse and the main pulse was fixed at 300~ps. In Appendix~\ref{ap:delay}, we present additional experimental data where the main pulse energy was fixed and the target held stationary near the peak focus, and where the pre-pulse energy and pre-pulse time delay were varied. This is an interesting parameter space to explore since varying the pre-pulse energy and timing will introduce different pre-plasma conditions to the target. Unfortunately, the data presented in Appendix~\ref{ap:delay} had significant target instability which hindered optimization efforts. The Appendix shows a proof of concept that time delay can be used in future optimization experiments with as many as 4 parameters. To the best of our knowledge, this is one of the largest data sets for ultra-intense laser acceleration of electrons and protons ever collected.

Another important aspect of our experiment was the ``liquid leaf" target system which is described well by \citet{George_etal2019}. A practical problem with these targets is that vibrations can lead to the breakup of the liquid sheet (e.g. \cite{liquidleaf}). Typically, studies with liquid leaf target systems involve relatively short timescales for data collection. For example, \citet{Treffert_2022} produced deuteron and electron spectra for 60 consecutive laser shots in an experiment lasting around 2 minutes. By comparison, the data we present here typically involved running a liquid target system for about a half hour. We note that one of the models (Ridge 4D) was likely affected by the target stability, but we believe the other results to be unaffected by target stability concerns.

During the optimization run, as highlighted in Figure~\ref{fig:beta_shift}, the number of electron and proton counts does increase significantly with decreasing $\beta$ while keeping $KE_{\rm cutoff,goal}$ fixed at 0.7~MeV as one would expect given the definition of Eq.~\ref{eq:function}.  For context, Figure~\ref{fig:beta_spectra} shows detailed results for one of the models as $\beta$ was varied. Efforts to shift the maximum energy of the electron energy distribution had limited success (Figure~\ref{fig:energy_shift}). The placement of the electron spectrometer likely had a strong impact on that result. As shown in Figure~\ref{fig:exp}, the electron spectrometer was placed in the target normal rear direction which raises the possibility that many electrons are being missed by the spectrometer. After these experiments were performed, and at a time when the liquid jet system was producing thin sheets with heavy water (as also used by \cite{Knight_etal2024}), we did briefly position the spectrometer in the forward laser direction.  This spectrometer location showed a significant increased the number of electron counts and the electron kinetic energies increased to about 0.9~MeV with some electrons over 1 MeV. We did not perform optimizations with this configuration so we omit these results from the current paper for brevity. It bears mentioning that the electron spectrometer has detected even higher electron energies in our lab previously. As described in \citet{Feister_etal2017}, when this same spectrometer was placed behind a hole in the OAP, electron energies were detected up to $\sim$2~MeV when the laser was near normal incidence on a liquid water column target. Clearly, there is interesting work to be done to explore different angles that electrons are ejected and which angles would be ideal for making a tunable MeV electron source. We left the spectrometer at target normal in this work because it is a dual electron/ion spectrometer and at these intensities the protons are being ejected from target normal.
    
To comment on the different models used, the ridge regression models performed better with higher dimensions, with the 8-dimensional model optimizing the best. The random forest optimized the electron spectra about equally well between 100 and 500 estimators. The Gaussian process regression optimized on-par with the other models. However, while other models optimized in seconds, the Gaussian process took tens of seconds to optimize. While this may not seem like a significant difference, it may prove important in future experiments as a more real-time optimization is implemented.

As mentioned earlier, the same experimental framework can be straightforwardly used to optimize and control proton acceleration. The ejected protons are likely more collimated than the electrons (c.f. single shot simulation results in \cite{Rahman_etal2021}), so a challenge with performing analogous experiments with protons is the proton spectrometer reading would be very sensitive to the stability of the liquid leaf target. In this experiment, the target suffered an instability that required adjustment 6 times. These instabilities are detrimental to the proton signal due to the signals narrow beam divergence. The electrons are accelerated in a much wider angle, which makes the electron signal more robust to the instabilities. Efforts are currently being made to improve the stability of the target to allow for proton optimization experiments.

Other radiation can potentially be optimized and controlled by these methods. Although we omit it for brevity, the electron/ion spectrometer in our experiment did detect x-rays which show up as counts in the parts of the line CCD that would indicate no deflection by the magnetic field. We could easily use this signal, which is connected to our high throughput DAQ system, in a scheme to increase or decrease the x-ray flux from the laser interaction region. And although the integration times would be longer, in principle one could optimize for neutron generation by using a heavy water or deuterated ethylene glycol liquid leaf target with the same suite of neutron detectors that were employed in an earlier paper from our group \cite{Knight_etal2024}.

\section{Conclusions}
\label{sec:concl}

We describe efforts to optimize and control MeV electrons ejected from ultra-intense laser interactions with a ``liquid leaf" target using a 10~mJ class, kHz repetition rate laser system at Wright Patterson AFB. MeV energy protons were also detected during these experiments which are well into the high repetition rate regime. Because of the high repetition rate,  significant experimental data was obtained and we used this data, which included over 50,000 electron energy spectra, to train different regression models in efforts to optimize and control the MeV electron signal. We also obtained over 5~million single-shot measurements of main pulse and pre-pulse energy. This work was greatly enabled by coupling some, but not all, of our instruments to EPICS protocol communications. Without completely redesigning the control system, we were able to perform sophisticated tests where algorithms directed the laser and target independent parameters. Many of the regression models we used were able to increase and decrease the electron number counts in a controlled way, but we were less successful in using these models shift the peak electron kinetic energy. In future work we plan to perform analogous experiments where the MeV proton signal is optimized, and potentially other sources of secondary radiation like x-rays and neutrons.

\begin{acknowledgments}
Supercomputer allocations for this project included time from the Ohio Supercomputer Center. We acknowledge support provided by the National Science Foundation (NSF) under Grant No. 2109222. Any opinions, findings, and conclusions or recommendations expressed in this material are those of the author(s) and do not necessarily reflect the views of the National Science Foundation. Dr. Orban and Dr. Snyder were both supported by the Air Force Office of Science Research summer faculty program. This work was supported by Air Force Office of Scientific Research (AFOSR) Award (PM: Dr. Andrew B. Stickrath). 
This work was also supported by Department of Energy (PM: Dr. Kramer Akli).
This paper has been cleared for public release, clearance number MSC/PA-2025-0006; 88ABW-2025-0018.
\end{acknowledgments}

\section*{Data Availability Statement}

The data that support the findings of this study are openly available zenodo\cite{tamminga_dataset} at \url{https://zenodo.org/records/15631826}, reference number 10.5281/zenodo.15631826

\clearpage
\appendix
\section{Time delay of the pre-pulse}

\label{ap:delay}

As discussed in main body of this paper, our primary experimental investigations assumed a time delay between the main pulse and the pre-pulse that was fixed to 300~ps. In principle, this pre-pulse time delay, which is controlled by a delay stage, could be one of the independent variables that is varied and used in an optimization scheme like the investigation described with the main-pulse energy, pre-pulse energy and the target position. We leave this for future work, but we here include results from an an additional parameter scan where the pre-pulse energy and pre-pulse delay time were varied while main pulse energy was held constant at its max level ($\sim$9~mJ) and the target held fixed near the peak focus.

\begin{figure*}
    \centering
    \includegraphics[width=\textwidth]{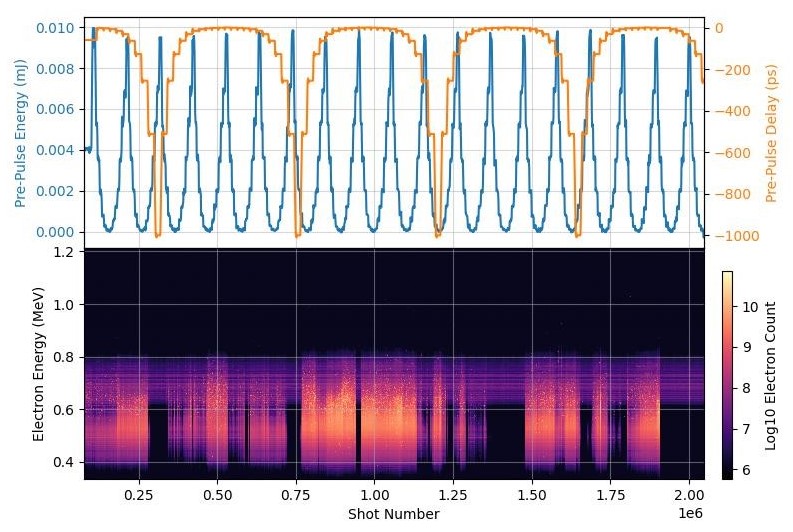}
    \caption{Details of a parameter scan which lasted 34 minutes and involved roughly 2.25 million laser shots. Upper panel: A plot showing the pre-pulse energy (blue) and pre-pulse delay (orange) as these quantities were varied over time. Bottom panel: A plot of the electron spectra collected during the parameter scan. The x-axis shows the shot number. There were no protons detected during this parameter scan. The spectra color bar is logarithmic with peak value corresponding to saturation of the CCD.}
    \label{fig:campaign2_param_scan}
\end{figure*}

This parameter scan lasted 40 minutes and it involved 2.4 million laser shots. Figure~\ref{fig:campaign2_param_scan} shows the varied parameters along with the corresponding electron spectra during the duration of the run. This parameter scan suffered from target instabilities which hindered the any optimization efforts using this dataset. Unlike the parameter scan described in Sec.~\ref{sec:results}, no significant proton signal was detected, so we do not include proton results in Figure.~\ref{fig:campaign2_param_scan}. As mentioned earlier, there is reason to believe that protons ejected from the laser interaction region are significantly more beamed than the electrons\cite{Rahman_etal2021}. So even a small misalignment between the spectrometer and the perpendicular to the liquid sheet can lead to the proton beam missing the entry port to the spectrometer. We presume that this is the reason why negligible protons were detected.

\begin{figure}
    \centering
    \includegraphics[width=\linewidth]{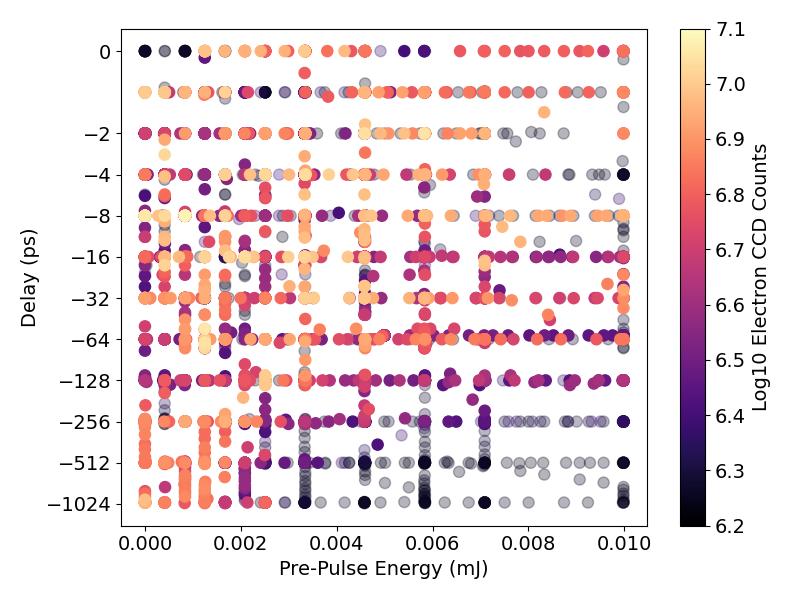}
    \caption{Pre-pulse energy and delay phase space explored during the parameter scan. A logarithmic color bar shows the total number of counts recorded by the electron spectrometer.}
    \label{fig:Param_space_scatter2}
\end{figure}

The phase space explored during this parameter scan can be seen in Figure \ref{fig:Param_space_scatter2}. One can see for 300~ps of time delay that the largest pre-pulse energies do significantly suppress the electron counts. Physically this is likely because of target expansion due to the pre-pulse heating. At higher than 300~ps of time delay, there is even more time for the target to expand and one sees a decrease in electron number counts even for lower pre-pulse energies. 

A notable result from this investigation (see Figure.~\ref{fig:campaign2_param_scan}) is that the pre-pulse time delay did affect the peak electron energies but at no point did the electron energies reach any higher than about 0.8~MeV, which is similar to what was found for other the other investigations in this paper. Figure~\ref{fig:Param_space_scatter2} and \ref{fig:campaign2_param_scan} present measurements where the electron spectrometer is placed at target normal. As mentioned earlier in Sec.~\ref{sec:discuss}, target normal may not be the best angle for the purpose of creating a tunable MeV electron source.

The results presented in this appendix represent a kind of proof of concept that time delay could be used in future optimization experiments. This prospect raises some interesting questions about what an efficient parameter scanning strategy would be when as many as four parameters (main pulse energy, pre-pulse energy, pre-pulse time delay and target position) can be varied, and how a parameter scan and optimization strategy can be designed to avoid local minima.

\clearpage
\bibliography{ms}

\begin{thebibliography}{29}
\expandafter\ifx\csname natexlab\endcsname\relax\def\natexlab#1{#1}\fi
\expandafter\ifx\csname bibnamefont\endcsname\relax
  \def\bibnamefont#1{#1}\fi
\expandafter\ifx\csname bibfnamefont\endcsname\relax
  \def\bibfnamefont#1{#1}\fi
\expandafter\ifx\csname citenamefont\endcsname\relax
  \def\citenamefont#1{#1}\fi
\expandafter\ifx\csname url\endcsname\relax
  \def\url#1{\texttt{#1}}\fi
\expandafter\ifx\csname urlprefix\endcsname\relax\def\urlprefix{URL }\fi
\providecommand{\bibinfo}[2]{#2}
\providecommand{\eprint}[2][]{\url{#2}}

\bibitem[{\citenamefont{Döpp et~al.}(2023)\citenamefont{Döpp, Eberle, Howard, Irshad, Lin, and Streeter}}]{dopp_etal2023}
\bibinfo{author}{\bibfnamefont{A.}~\bibnamefont{Döpp}}, \bibinfo{author}{\bibfnamefont{C.}~\bibnamefont{Eberle}}, \bibinfo{author}{\bibfnamefont{S.}~\bibnamefont{Howard}}, \bibinfo{author}{\bibfnamefont{F.}~\bibnamefont{Irshad}}, \bibinfo{author}{\bibfnamefont{J.}~\bibnamefont{Lin}}, \bibnamefont{and} \bibinfo{author}{\bibfnamefont{M.}~\bibnamefont{Streeter}}, \bibinfo{journal}{High Power Laser Science and Engineering} p. \bibinfo{pages}{1–50} (\bibinfo{year}{2023}).

\bibitem[{\citenamefont{Anirudh et~al.}(2022)\citenamefont{Anirudh, Archibald, Asif, Becker, Benkadda, Bremer, Budé, Chang, Chen, Churchill et~al.}}]{Anirudh_etal2022}
\bibinfo{author}{\bibfnamefont{R.}~\bibnamefont{Anirudh}}, \bibinfo{author}{\bibfnamefont{R.}~\bibnamefont{Archibald}}, \bibinfo{author}{\bibfnamefont{M.~S.} \bibnamefont{Asif}}, \bibinfo{author}{\bibfnamefont{M.~M.} \bibnamefont{Becker}}, \bibinfo{author}{\bibfnamefont{S.}~\bibnamefont{Benkadda}}, \bibinfo{author}{\bibfnamefont{P.-T.} \bibnamefont{Bremer}}, \bibinfo{author}{\bibfnamefont{R.~H.~S.} \bibnamefont{Budé}}, \bibinfo{author}{\bibfnamefont{C.~S.} \bibnamefont{Chang}}, \bibinfo{author}{\bibfnamefont{L.}~\bibnamefont{Chen}}, \bibinfo{author}{\bibfnamefont{R.~M.} \bibnamefont{Churchill}}, \bibnamefont{et~al.}, \emph{\bibinfo{title}{2022 review of data-driven plasma science}} (\bibinfo{year}{2022}), \eprint{2205.15832}.

\bibitem[{\citenamefont{He et~al.}(2015)\citenamefont{He, Hou, Gao, Lebailly, Nees, Clarke, Krushelnick, and Thomas}}]{He_etal2015}
\bibinfo{author}{\bibfnamefont{Z.-H.} \bibnamefont{He}}, \bibinfo{author}{\bibfnamefont{B.}~\bibnamefont{Hou}}, \bibinfo{author}{\bibfnamefont{G.}~\bibnamefont{Gao}}, \bibinfo{author}{\bibfnamefont{V.}~\bibnamefont{Lebailly}}, \bibinfo{author}{\bibfnamefont{J.~A.} \bibnamefont{Nees}}, \bibinfo{author}{\bibfnamefont{R.}~\bibnamefont{Clarke}}, \bibinfo{author}{\bibfnamefont{K.}~\bibnamefont{Krushelnick}}, \bibnamefont{and} \bibinfo{author}{\bibfnamefont{A.~G.~R.} \bibnamefont{Thomas}}, \bibinfo{journal}{Physics of Plasmas} \textbf{\bibinfo{volume}{22}}, \bibinfo{pages}{056704} (\bibinfo{year}{2015}), ISSN \bibinfo{issn}{1070-664X}, \eprint{https://pubs.aip.org/aip/pop/article-pdf/doi/10.1063/1.4921159/13789642/056704\_1\_online.pdf}, \urlprefix\url{https://doi.org/10.1063/1.4921159}.

\bibitem[{\citenamefont{Lin et~al.}(2021)\citenamefont{Lin, Qian, Murphy, Hsu, Hero, Ma, Thomas, and Krushelnick}}]{Lin_etal2021}
\bibinfo{author}{\bibfnamefont{J.}~\bibnamefont{Lin}}, \bibinfo{author}{\bibfnamefont{Q.}~\bibnamefont{Qian}}, \bibinfo{author}{\bibfnamefont{J.}~\bibnamefont{Murphy}}, \bibinfo{author}{\bibfnamefont{A.}~\bibnamefont{Hsu}}, \bibinfo{author}{\bibfnamefont{A.}~\bibnamefont{Hero}}, \bibinfo{author}{\bibfnamefont{Y.}~\bibnamefont{Ma}}, \bibinfo{author}{\bibfnamefont{A.~G.~R.} \bibnamefont{Thomas}}, \bibnamefont{and} \bibinfo{author}{\bibfnamefont{K.}~\bibnamefont{Krushelnick}}, \bibinfo{journal}{Physics of Plasmas} \textbf{\bibinfo{volume}{28}}, \bibinfo{pages}{083102} (\bibinfo{year}{2021}), ISSN \bibinfo{issn}{1070-664X}, \eprint{https://pubs.aip.org/aip/pop/article-pdf/doi/10.1063/5.0047940/13648447/083102\_1\_online.pdf}, \urlprefix\url{https://doi.org/10.1063/5.0047940}.

\bibitem[{\citenamefont{Jalas et~al.}(2021)\citenamefont{Jalas, Kirchen, Messner, Winkler, H\"ubner, Dirkwinkel, Schnepp, Lehe, and Maier}}]{Jalas_etal2021}
\bibinfo{author}{\bibfnamefont{S.}~\bibnamefont{Jalas}}, \bibinfo{author}{\bibfnamefont{M.}~\bibnamefont{Kirchen}}, \bibinfo{author}{\bibfnamefont{P.}~\bibnamefont{Messner}}, \bibinfo{author}{\bibfnamefont{P.}~\bibnamefont{Winkler}}, \bibinfo{author}{\bibfnamefont{L.}~\bibnamefont{H\"ubner}}, \bibinfo{author}{\bibfnamefont{J.}~\bibnamefont{Dirkwinkel}}, \bibinfo{author}{\bibfnamefont{M.}~\bibnamefont{Schnepp}}, \bibinfo{author}{\bibfnamefont{R.}~\bibnamefont{Lehe}}, \bibnamefont{and} \bibinfo{author}{\bibfnamefont{A.~R.} \bibnamefont{Maier}}, \bibinfo{journal}{Phys. Rev. Lett.} \textbf{\bibinfo{volume}{126}}, \bibinfo{pages}{104801} (\bibinfo{year}{2021}), \urlprefix\url{https://link.aps.org/doi/10.1103/PhysRevLett.126.104801}.

\bibitem[{\citenamefont{Ye et~al.}(2022)\citenamefont{Ye, Gu, Zhang, Wang, Tan, Zhang, Yang, Yan, Wu, Huang et~al.}}]{Ye_etal2022}
\bibinfo{author}{\bibfnamefont{H.}~\bibnamefont{Ye}}, \bibinfo{author}{\bibfnamefont{Y.}~\bibnamefont{Gu}}, \bibinfo{author}{\bibfnamefont{X.}~\bibnamefont{Zhang}}, \bibinfo{author}{\bibfnamefont{S.}~\bibnamefont{Wang}}, \bibinfo{author}{\bibfnamefont{F.}~\bibnamefont{Tan}}, \bibinfo{author}{\bibfnamefont{J.}~\bibnamefont{Zhang}}, \bibinfo{author}{\bibfnamefont{Y.}~\bibnamefont{Yang}}, \bibinfo{author}{\bibfnamefont{Y.}~\bibnamefont{Yan}}, \bibinfo{author}{\bibfnamefont{Y.}~\bibnamefont{Wu}}, \bibinfo{author}{\bibfnamefont{W.}~\bibnamefont{Huang}}, \bibnamefont{et~al.}, \bibinfo{journal}{Results in Physics} \textbf{\bibinfo{volume}{43}}, \bibinfo{pages}{106116} (\bibinfo{year}{2022}), ISSN \bibinfo{issn}{2211-3797}, \urlprefix\url{https://www.sciencedirect.com/science/article/pii/S2211379722007306}.

\bibitem[{\citenamefont{Dann et~al.}(2019)\citenamefont{Dann, Baird, Bourgeois, Chekhlov, Eardley, Gregory, Gruse, Hah, Hazra, Hawkes et~al.}}]{Dann_etal2019}
\bibinfo{author}{\bibfnamefont{S.~J.~D.} \bibnamefont{Dann}}, \bibinfo{author}{\bibfnamefont{C.~D.} \bibnamefont{Baird}}, \bibinfo{author}{\bibfnamefont{N.}~\bibnamefont{Bourgeois}}, \bibinfo{author}{\bibfnamefont{O.}~\bibnamefont{Chekhlov}}, \bibinfo{author}{\bibfnamefont{S.}~\bibnamefont{Eardley}}, \bibinfo{author}{\bibfnamefont{C.~D.} \bibnamefont{Gregory}}, \bibinfo{author}{\bibfnamefont{J.-N.} \bibnamefont{Gruse}}, \bibinfo{author}{\bibfnamefont{J.}~\bibnamefont{Hah}}, \bibinfo{author}{\bibfnamefont{D.}~\bibnamefont{Hazra}}, \bibinfo{author}{\bibfnamefont{S.~J.} \bibnamefont{Hawkes}}, \bibnamefont{et~al.}, \bibinfo{journal}{Phys. Rev. Accel. Beams} \textbf{\bibinfo{volume}{22}}, \bibinfo{pages}{041303} (\bibinfo{year}{2019}), \urlprefix\url{https://link.aps.org/doi/10.1103/PhysRevAccelBeams.22.041303}.

\bibitem[{\citenamefont{Mariscal et~al.}(2024)\citenamefont{Mariscal, Djordjevic, Anirudh, Jayaraman-Thiagarajan, Grace, Simpson, Swanson, Galvin, Mittelberger, Heebner et~al.}}]{Mariscal_etal2024}
\bibinfo{author}{\bibfnamefont{D.~A.} \bibnamefont{Mariscal}}, \bibinfo{author}{\bibfnamefont{B.~Z.} \bibnamefont{Djordjevic}}, \bibinfo{author}{\bibfnamefont{R.}~\bibnamefont{Anirudh}}, \bibinfo{author}{\bibfnamefont{J.}~\bibnamefont{Jayaraman-Thiagarajan}}, \bibinfo{author}{\bibfnamefont{E.~S.} \bibnamefont{Grace}}, \bibinfo{author}{\bibfnamefont{R.~A.} \bibnamefont{Simpson}}, \bibinfo{author}{\bibfnamefont{K.~K.} \bibnamefont{Swanson}}, \bibinfo{author}{\bibfnamefont{T.~C.} \bibnamefont{Galvin}}, \bibinfo{author}{\bibfnamefont{D.}~\bibnamefont{Mittelberger}}, \bibinfo{author}{\bibfnamefont{J.~E.} \bibnamefont{Heebner}}, \bibnamefont{et~al.}, \bibinfo{journal}{Physics of Plasmas} \textbf{\bibinfo{volume}{31}}, \bibinfo{pages}{073105} (\bibinfo{year}{2024}), ISSN \bibinfo{issn}{1070-664X}, \eprint{https://pubs.aip.org/aip/pop/article-pdf/doi/10.1063/5.0190553/20050018/073105\_1\_5.0190553.pdf}, \urlprefix\url{https://doi.org/10.1063/5.0190553}.

\bibitem[{\citenamefont{ul~Haq et~al.}(2018)\citenamefont{ul~Haq, Ahmed, Sokollik, Fang, Ge, Yuan, and Chen}}]{Noaman_etal2018}
\bibinfo{author}{\bibfnamefont{M.~N.} \bibnamefont{ul~Haq}}, \bibinfo{author}{\bibfnamefont{H.}~\bibnamefont{Ahmed}}, \bibinfo{author}{\bibfnamefont{T.}~\bibnamefont{Sokollik}}, \bibinfo{author}{\bibfnamefont{Y.}~\bibnamefont{Fang}}, \bibinfo{author}{\bibfnamefont{X.}~\bibnamefont{Ge}}, \bibinfo{author}{\bibfnamefont{X.}~\bibnamefont{Yuan}}, \bibnamefont{and} \bibinfo{author}{\bibfnamefont{L.}~\bibnamefont{Chen}}, \bibinfo{journal}{Nuclear Instruments and Methods in Physics Research Section A: Accelerators, Spectrometers, Detectors and Associated Equipment} \textbf{\bibinfo{volume}{909}}, \bibinfo{pages}{164} (\bibinfo{year}{2018}), ISSN \bibinfo{issn}{0168-9002}, \bibinfo{note}{3rd European Advanced Accelerator Concepts workshop (EAAC2017)}, \urlprefix\url{https://www.sciencedirect.com/science/article/pii/S0168900218302961}.

\bibitem[{\citenamefont{{Ma} et~al.}(2021)\citenamefont{{Ma}, {Mariscal}, {Anirudh}, {Bremer}, {Djordjevic}, {Galvin}, {Grace}, {Herriot}, {Jacobs}, {Kailkhura} et~al.}}]{Ma_etal2021}
\bibinfo{author}{\bibfnamefont{T.}~\bibnamefont{{Ma}}}, \bibinfo{author}{\bibfnamefont{D.}~\bibnamefont{{Mariscal}}}, \bibinfo{author}{\bibfnamefont{R.}~\bibnamefont{{Anirudh}}}, \bibinfo{author}{\bibfnamefont{T.}~\bibnamefont{{Bremer}}}, \bibinfo{author}{\bibfnamefont{B.~Z.} \bibnamefont{{Djordjevic}}}, \bibinfo{author}{\bibfnamefont{T.}~\bibnamefont{{Galvin}}}, \bibinfo{author}{\bibfnamefont{E.}~\bibnamefont{{Grace}}}, \bibinfo{author}{\bibfnamefont{S.}~\bibnamefont{{Herriot}}}, \bibinfo{author}{\bibfnamefont{S.}~\bibnamefont{{Jacobs}}}, \bibinfo{author}{\bibfnamefont{B.}~\bibnamefont{{Kailkhura}}}, \bibnamefont{et~al.}, \bibinfo{journal}{Plasma Physics and Controlled Fusion} \textbf{\bibinfo{volume}{63}}, \bibinfo{eid}{104003} (\bibinfo{year}{2021}).

\bibitem[{\citenamefont{Palmer}(2018)}]{Palmer2018}
\bibinfo{author}{\bibfnamefont{C.}~\bibnamefont{Palmer}}, \bibinfo{journal}{New Journal of Physics} \textbf{\bibinfo{volume}{20}}, \bibinfo{pages}{061001} (\bibinfo{year}{2018}), \urlprefix\url{https://dx.doi.org/10.1088/1367-2630/aac5ce}.

\bibitem[{\citenamefont{{Hooker}}(2013)}]{Hooker2013}
\bibinfo{author}{\bibfnamefont{S.~M.} \bibnamefont{{Hooker}}}, \bibinfo{journal}{Nature Photonics} \textbf{\bibinfo{volume}{7}}, \bibinfo{pages}{775} (\bibinfo{year}{2013}), \eprint{1406.5118}.

\bibitem[{\citenamefont{Heuer et~al.}(2022)\citenamefont{Heuer, Feister, Schaeffer, and Rinderknecht}}]{Heuer_etal2022}
\bibinfo{author}{\bibfnamefont{P.~V.} \bibnamefont{Heuer}}, \bibinfo{author}{\bibfnamefont{S.}~\bibnamefont{Feister}}, \bibinfo{author}{\bibfnamefont{D.~B.} \bibnamefont{Schaeffer}}, \bibnamefont{and} \bibinfo{author}{\bibfnamefont{H.~G.} \bibnamefont{Rinderknecht}}, \bibinfo{journal}{Physics of Plasmas} \textbf{\bibinfo{volume}{29}} (\bibinfo{year}{2022}), ISSN \bibinfo{issn}{1070-664X}, \bibinfo{note}{110401}, \eprint{https://pubs.aip.org/aip/pop/article-pdf/doi/10.1063/5.0130801/16624799/110401\_1\_online.pdf}, \urlprefix\url{https://doi.org/10.1063/5.0130801}.

\bibitem[{\citenamefont{Feister et~al.}(2023)\citenamefont{Feister, Cassou, Dann, Döpp, Gauron, Gonsalves, Joglekar, Marshall, Neveu, Schlenvoigt et~al.}}]{Feister_etal2023}
\bibinfo{author}{\bibfnamefont{S.}~\bibnamefont{Feister}}, \bibinfo{author}{\bibfnamefont{K.}~\bibnamefont{Cassou}}, \bibinfo{author}{\bibfnamefont{S.}~\bibnamefont{Dann}}, \bibinfo{author}{\bibfnamefont{A.}~\bibnamefont{Döpp}}, \bibinfo{author}{\bibfnamefont{P.}~\bibnamefont{Gauron}}, \bibinfo{author}{\bibfnamefont{A.~J.} \bibnamefont{Gonsalves}}, \bibinfo{author}{\bibfnamefont{A.}~\bibnamefont{Joglekar}}, \bibinfo{author}{\bibfnamefont{V.}~\bibnamefont{Marshall}}, \bibinfo{author}{\bibfnamefont{O.}~\bibnamefont{Neveu}}, \bibinfo{author}{\bibfnamefont{H.-P.} \bibnamefont{Schlenvoigt}}, \bibnamefont{et~al.}, \bibinfo{journal}{High Power Laser Science and Engineering} \textbf{\bibinfo{volume}{11}}, \bibinfo{pages}{e56} (\bibinfo{year}{2023}).

\bibitem[{\citenamefont{{Morrison} et~al.}(2018)\citenamefont{{Morrison}, {Feister}, {Frische}, {Austin}, {Ngirmang}, {Murphy}, {Orban}, {Chowdhury}, and {Roquemore}}}]{Morrison_etal2018}
\bibinfo{author}{\bibfnamefont{J.~T.} \bibnamefont{{Morrison}}}, \bibinfo{author}{\bibfnamefont{S.}~\bibnamefont{{Feister}}}, \bibinfo{author}{\bibfnamefont{K.~D.} \bibnamefont{{Frische}}}, \bibinfo{author}{\bibfnamefont{D.~R.} \bibnamefont{{Austin}}}, \bibinfo{author}{\bibfnamefont{G.~K.} \bibnamefont{{Ngirmang}}}, \bibinfo{author}{\bibfnamefont{N.~R.} \bibnamefont{{Murphy}}}, \bibinfo{author}{\bibfnamefont{C.}~\bibnamefont{{Orban}}}, \bibinfo{author}{\bibfnamefont{E.~A.} \bibnamefont{{Chowdhury}}}, \bibnamefont{and} \bibinfo{author}{\bibfnamefont{W.~M.} \bibnamefont{{Roquemore}}}, \bibinfo{journal}{New Journal of Physics} \textbf{\bibinfo{volume}{20}}, \bibinfo{eid}{022001} (\bibinfo{year}{2018}).

\bibitem[{\citenamefont{Feister et~al.}(2017)\citenamefont{Feister, Austin, Morrison, Frische, Orban, Ngirmang, Handler, Smith, Schillaci, LaVerne et~al.}}]{Feister_etal2017}
\bibinfo{author}{\bibfnamefont{S.}~\bibnamefont{Feister}}, \bibinfo{author}{\bibfnamefont{D.~R.} \bibnamefont{Austin}}, \bibinfo{author}{\bibfnamefont{J.~T.} \bibnamefont{Morrison}}, \bibinfo{author}{\bibfnamefont{K.~D.} \bibnamefont{Frische}}, \bibinfo{author}{\bibfnamefont{C.}~\bibnamefont{Orban}}, \bibinfo{author}{\bibfnamefont{G.}~\bibnamefont{Ngirmang}}, \bibinfo{author}{\bibfnamefont{A.}~\bibnamefont{Handler}}, \bibinfo{author}{\bibfnamefont{J.~R.~H.} \bibnamefont{Smith}}, \bibinfo{author}{\bibfnamefont{M.}~\bibnamefont{Schillaci}}, \bibinfo{author}{\bibfnamefont{J.~A.} \bibnamefont{LaVerne}}, \bibnamefont{et~al.}, \bibinfo{journal}{Opt. Express} \textbf{\bibinfo{volume}{25}}, \bibinfo{pages}{18736} (\bibinfo{year}{2017}), \urlprefix\url{https://opg.optica.org/oe/abstract.cfm?URI=oe-25-16-18736}.

\bibitem[{\citenamefont{Morrison et~al.}(2015)\citenamefont{Morrison, Chowdhury, Frische, Feister, Ovchinnikov, Nees, Orban, Freeman, and Roquemore}}]{Morrison_etal2015}
\bibinfo{author}{\bibfnamefont{J.~T.} \bibnamefont{Morrison}}, \bibinfo{author}{\bibfnamefont{E.~A.} \bibnamefont{Chowdhury}}, \bibinfo{author}{\bibfnamefont{K.~D.} \bibnamefont{Frische}}, \bibinfo{author}{\bibfnamefont{S.}~\bibnamefont{Feister}}, \bibinfo{author}{\bibfnamefont{V.~M.} \bibnamefont{Ovchinnikov}}, \bibinfo{author}{\bibfnamefont{J.~A.} \bibnamefont{Nees}}, \bibinfo{author}{\bibfnamefont{C.}~\bibnamefont{Orban}}, \bibinfo{author}{\bibfnamefont{R.~R.} \bibnamefont{Freeman}}, \bibnamefont{and} \bibinfo{author}{\bibfnamefont{W.~M.} \bibnamefont{Roquemore}}, \bibinfo{journal}{Physics of Plasmas} \textbf{\bibinfo{volume}{22}}, \bibinfo{pages}{043101} (\bibinfo{year}{2015}), ISSN \bibinfo{issn}{1070-664X}, \eprint{https://pubs.aip.org/aip/pop/article-pdf/doi/10.1063/1.4916493/14018848/043101\_1\_online.pdf}, \urlprefix\url{https://doi.org/10.1063/1.4916493}.

\bibitem[{\citenamefont{Knight et~al.}(2024)\citenamefont{Knight, Gautam, Stoner, Egner, Smith, Orban, Manfredi, Frische, Dexter, Chowdhury et~al.}}]{Knight_etal2024}
\bibinfo{author}{\bibfnamefont{B.~M.} \bibnamefont{Knight}}, \bibinfo{author}{\bibfnamefont{C.~M.} \bibnamefont{Gautam}}, \bibinfo{author}{\bibfnamefont{C.~R.} \bibnamefont{Stoner}}, \bibinfo{author}{\bibfnamefont{B.~V.} \bibnamefont{Egner}}, \bibinfo{author}{\bibfnamefont{J.~R.} \bibnamefont{Smith}}, \bibinfo{author}{\bibfnamefont{C.~M.} \bibnamefont{Orban}}, \bibinfo{author}{\bibfnamefont{J.~J.} \bibnamefont{Manfredi}}, \bibinfo{author}{\bibfnamefont{K.~D.} \bibnamefont{Frische}}, \bibinfo{author}{\bibfnamefont{M.~L.} \bibnamefont{Dexter}}, \bibinfo{author}{\bibfnamefont{E.~A.} \bibnamefont{Chowdhury}}, \bibnamefont{et~al.}, \bibinfo{journal}{High Power Laser Science and Engineering} \textbf{\bibinfo{volume}{12}}, \bibinfo{pages}{e2} (\bibinfo{year}{2024}).

\bibitem[{\citenamefont{Desai et~al.}(2024)\citenamefont{Desai, Zhang, Oropeza, Felice, Smith, Kryshchenko, Orban, Dexter, and Patnaik}}]{Desai_etal2024}
\bibinfo{author}{\bibfnamefont{R.}~\bibnamefont{Desai}}, \bibinfo{author}{\bibfnamefont{T.}~\bibnamefont{Zhang}}, \bibinfo{author}{\bibfnamefont{R.}~\bibnamefont{Oropeza}}, \bibinfo{author}{\bibfnamefont{J.~J.} \bibnamefont{Felice}}, \bibinfo{author}{\bibfnamefont{J.~R.} \bibnamefont{Smith}}, \bibinfo{author}{\bibfnamefont{A.}~\bibnamefont{Kryshchenko}}, \bibinfo{author}{\bibfnamefont{C.}~\bibnamefont{Orban}}, \bibinfo{author}{\bibfnamefont{M.~L.} \bibnamefont{Dexter}}, \bibnamefont{and} \bibinfo{author}{\bibfnamefont{A.~K.} \bibnamefont{Patnaik}}, \emph{\bibinfo{title}{Applying machine learning methods to laser acceleration of protons: Lessons learned from synthetic data}} (\bibinfo{year}{2024}), \eprint{2307.16036}, \urlprefix\url{https://arxiv.org/abs/2307.16036}.

\bibitem[{\citenamefont{Feister et~al.}(2014)\citenamefont{Feister, Nees, Morrison, Frische, Orban, Chowdhury, and Roquemore}}]{Feister_etal2014}
\bibinfo{author}{\bibfnamefont{S.}~\bibnamefont{Feister}}, \bibinfo{author}{\bibfnamefont{J.~A.} \bibnamefont{Nees}}, \bibinfo{author}{\bibfnamefont{J.~T.} \bibnamefont{Morrison}}, \bibinfo{author}{\bibfnamefont{K.~D.} \bibnamefont{Frische}}, \bibinfo{author}{\bibfnamefont{C.}~\bibnamefont{Orban}}, \bibinfo{author}{\bibfnamefont{E.~A.} \bibnamefont{Chowdhury}}, \bibnamefont{and} \bibinfo{author}{\bibfnamefont{W.~M.} \bibnamefont{Roquemore}}, \bibinfo{journal}{Review of Scientific Instruments} \textbf{\bibinfo{volume}{85}}, \bibinfo{pages}{11D602} (\bibinfo{year}{2014}), ISSN \bibinfo{issn}{0034-6748}, \eprint{https://pubs.aip.org/aip/rsi/article-pdf/doi/10.1063/1.4886955/13418446/11d602\_1\_online.pdf}, \urlprefix\url{https://doi.org/10.1063/1.4886955}.

\bibitem[{\citenamefont{George et~al.}(2019)\citenamefont{George, Morrison, Feister, Ngirmang, Smith, Klim, Snyder, Austin, Erbsen, Frische et~al.}}]{George_etal2019}
\bibinfo{author}{\bibfnamefont{K.~M.} \bibnamefont{George}}, \bibinfo{author}{\bibfnamefont{J.~T.} \bibnamefont{Morrison}}, \bibinfo{author}{\bibfnamefont{S.}~\bibnamefont{Feister}}, \bibinfo{author}{\bibfnamefont{G.~K.} \bibnamefont{Ngirmang}}, \bibinfo{author}{\bibfnamefont{J.~R.} \bibnamefont{Smith}}, \bibinfo{author}{\bibfnamefont{A.~J.} \bibnamefont{Klim}}, \bibinfo{author}{\bibfnamefont{J.}~\bibnamefont{Snyder}}, \bibinfo{author}{\bibfnamefont{D.}~\bibnamefont{Austin}}, \bibinfo{author}{\bibfnamefont{W.}~\bibnamefont{Erbsen}}, \bibinfo{author}{\bibfnamefont{K.~D.} \bibnamefont{Frische}}, \bibnamefont{et~al.}, \bibinfo{journal}{High Power Laser Science and Engineering} \textbf{\bibinfo{volume}{7}}, \bibinfo{pages}{e50} (\bibinfo{year}{2019}).

\bibitem[{\citenamefont{Ge et~al.}(2017)\citenamefont{Ge, Yang, and Han}}]{Ge_etal2017}
\bibinfo{author}{\bibfnamefont{X.}~\bibnamefont{Ge}}, \bibinfo{author}{\bibfnamefont{F.}~\bibnamefont{Yang}}, \bibnamefont{and} \bibinfo{author}{\bibfnamefont{Q.-L.} \bibnamefont{Han}}, \bibinfo{journal}{Information Sciences} \textbf{\bibinfo{volume}{380}}, \bibinfo{pages}{117} (\bibinfo{year}{2017}), ISSN \bibinfo{issn}{0020-0255}, \urlprefix\url{https://www.sciencedirect.com/science/article/pii/S0020025515005551}.

\bibitem[{\citenamefont{Institute}(2024)}]{streamdevice_2024}
\bibinfo{author}{\bibfnamefont{P.~S.} \bibnamefont{Institute}}, \emph{\bibinfo{title}{Streamdevice}}, \bibinfo{howpublished}{\url{https://paulscherrerinstitute.github.io/StreamDevice/}} (\bibinfo{year}{2024}), \bibinfo{note}{accessed: 2024-09-20}.

\bibitem[{\citenamefont{Extensions}(2024)}]{calab_2024}
\bibinfo{author}{\bibfnamefont{E.}~\bibnamefont{Extensions}}, \emph{\bibinfo{title}{Calab}}, \bibinfo{howpublished}{\url{https://github.com/epics-extensions/CALab}} (\bibinfo{year}{2024}), \bibinfo{note}{accessed: 2024-09-20}.

\bibitem[{\citenamefont{Pedregosa et~al.}(2011)\citenamefont{Pedregosa, Varoquaux, Gramfort, Michel, Thirion, Grisel, Blondel, Prettenhofer, Weiss, Dubourg et~al.}}]{scikit-learn}
\bibinfo{author}{\bibfnamefont{F.}~\bibnamefont{Pedregosa}}, \bibinfo{author}{\bibfnamefont{G.}~\bibnamefont{Varoquaux}}, \bibinfo{author}{\bibfnamefont{A.}~\bibnamefont{Gramfort}}, \bibinfo{author}{\bibfnamefont{V.}~\bibnamefont{Michel}}, \bibinfo{author}{\bibfnamefont{B.}~\bibnamefont{Thirion}}, \bibinfo{author}{\bibfnamefont{O.}~\bibnamefont{Grisel}}, \bibinfo{author}{\bibfnamefont{M.}~\bibnamefont{Blondel}}, \bibinfo{author}{\bibfnamefont{P.}~\bibnamefont{Prettenhofer}}, \bibinfo{author}{\bibfnamefont{R.}~\bibnamefont{Weiss}}, \bibinfo{author}{\bibfnamefont{V.}~\bibnamefont{Dubourg}}, \bibnamefont{et~al.}, \bibinfo{journal}{Journal of Machine Learning Research} \textbf{\bibinfo{volume}{12}}, \bibinfo{pages}{2825} (\bibinfo{year}{2011}), \urlprefix\url{https://scikit-learn.org/stable/index.html}.

\bibitem[{\citenamefont{Margarone et~al.}(2023)\citenamefont{Margarone, Schmitz, Kreuter, and Boine-Frankenheim}}]{liquidleaf}
\bibinfo{author}{\bibfnamefont{D.}~\bibnamefont{Margarone}}, \bibinfo{author}{\bibfnamefont{B.}~\bibnamefont{Schmitz}}, \bibinfo{author}{\bibfnamefont{D.}~\bibnamefont{Kreuter}}, \bibnamefont{and} \bibinfo{author}{\bibfnamefont{O.}~\bibnamefont{Boine-Frankenheim}}, \bibinfo{journal}{Laser and Particle Beams} \textbf{\bibinfo{volume}{2023}}, \bibinfo{pages}{2868112} (\bibinfo{year}{2023}), ISSN \bibinfo{issn}{0263-0346}, \urlprefix\url{https://doi.org/10.1155/2023/2868112}.

\bibitem[{\citenamefont{Treffert et~al.}(2022)\citenamefont{Treffert, Curry, Chou, Crissman, DePonte, Fiuza, Glenn, Hollinger, Nedbailo, Park et~al.}}]{Treffert_2022}
\bibinfo{author}{\bibfnamefont{F.}~\bibnamefont{Treffert}}, \bibinfo{author}{\bibfnamefont{C.~B.} \bibnamefont{Curry}}, \bibinfo{author}{\bibfnamefont{H.-G.~J.} \bibnamefont{Chou}}, \bibinfo{author}{\bibfnamefont{C.~J.} \bibnamefont{Crissman}}, \bibinfo{author}{\bibfnamefont{D.~P.} \bibnamefont{DePonte}}, \bibinfo{author}{\bibfnamefont{F.}~\bibnamefont{Fiuza}}, \bibinfo{author}{\bibfnamefont{G.~D.} \bibnamefont{Glenn}}, \bibinfo{author}{\bibfnamefont{R.~C.} \bibnamefont{Hollinger}}, \bibinfo{author}{\bibfnamefont{R.}~\bibnamefont{Nedbailo}}, \bibinfo{author}{\bibfnamefont{J.}~\bibnamefont{Park}}, \bibnamefont{et~al.}, \bibinfo{journal}{Applied Physics Letters} \textbf{\bibinfo{volume}{121}}, \bibinfo{pages}{074104} (\bibinfo{year}{2022}), ISSN \bibinfo{issn}{0003-6951}, \eprint{https://pubs.aip.org/aip/apl/article-pdf/doi/10.1063/5.0098973/16484736/074104\_1\_online.pdf}, \urlprefix\url{https://doi.org/10.1063/5.0098973}.

\bibitem[{\citenamefont{Rahman et~al.}(2021)\citenamefont{Rahman, Smith, Ngirmang, and Orban}}]{Rahman_etal2021}
\bibinfo{author}{\bibfnamefont{N.}~\bibnamefont{Rahman}}, \bibinfo{author}{\bibfnamefont{J.~R.} \bibnamefont{Smith}}, \bibinfo{author}{\bibfnamefont{G.~K.} \bibnamefont{Ngirmang}}, \bibnamefont{and} \bibinfo{author}{\bibfnamefont{C.}~\bibnamefont{Orban}}, \bibinfo{journal}{Physics of Plasmas} \textbf{\bibinfo{volume}{28}}, \bibinfo{pages}{073103} (\bibinfo{year}{2021}), ISSN \bibinfo{issn}{1070-664X}, \eprint{https://pubs.aip.org/aip/pop/article-pdf/doi/10.1063/5.0045320/16067211/073103\_1\_online.pdf}, \urlprefix\url{https://doi.org/10.1063/5.0045320}.

\bibitem[{\citenamefont{Tamminga}(2025)}]{tamminga_dataset}
\bibinfo{author}{\bibfnamefont{N.}~\bibnamefont{Tamminga}}, \emph{\bibinfo{title}{Towards intelligent control of mev electrons and protons from khz repetition rate ultra-intense laser interactions [data set]}}, \bibinfo{howpublished}{\url{https://doi.org/10.5281/zenodo.15631826}} (\bibinfo{year}{2025}).

\end{thebibliography}
\bibliographystyle{apsrev}

\end{document}